\documentstyle[prd,aps,eqsecnum,preprint,tighten,floats]{revtex}
\def\del{\delta}

\begin{document}
\draft
\preprint{\vbox{\hfill OHSTPY--HEP--TH--98--004\\
          \vbox{\vskip0.5in}}
         }

\title{Bound States of Dimensionally Reduced $\mbox{SYM}_{2+1}$ at Finite N }

\author{Francesco Antonuccio, Oleg Lunin, Stephen S. Pinsky}

\address{Department of Physics,\\ The Ohio State University,\\ Columbus,
OH 43210}

\date{\today}

\maketitle

\begin{abstract}
We consider the dimensional reduction of ${\cal N} = 1$
$\mbox{SYM}_{2+1}$ to $1+1$ dimensions. The gauge groups
we consider are U($N$) and SU($N$), where $N$ is finite.
We formulate the continuum bound state problem
in the light-cone formalism, and show that 
any normalizable SU($N$) bound state  must be a superposition
of an infinite number of Fock states. 
We also discuss how massless states arise in the DLCQ formulation
for certain discretizations.
\end{abstract}
\pacs{ }

\begin{sloppypar}
\section{ \bf Introduction.}
\renewcommand{\theequation}{1.\arabic{equation}}
\setcounter{equation}{0}

Solving for the non-perturbative properties of 
quantum field theories -- such as QCD -- is typically
an intractable problem. In order to gain some  
insights, however, a number of lower
dimensional models have been investigated in the large $N$ (planar) 
approximation, with a plethora of examples emerging in the
last few years (for a review see\cite{bpp98}). 

In this work, we will pursue the same theme of studying
a lower dimensional field theory, but unlike many previous 
investigations, we will allow the number of gauge colors, $N$, to be finite.
In particular, we will consider the $1+1$ dimensional theory that is 
obtained by dimensionally reducing $2+1$ dimensional
${\cal N} = 1$ supersymmetric Yang Mills theory.
The large $N$ limit
of this theory has already been
investigated in \cite{mss95}, and was shown to exhibit
the phenomena of screening \cite{gkm96,ars97}. 
In this work, we will find it advantageous to quantize the theory
on the light-cone, and to adopt the light-cone gauge.
Since the light-cone Hamiltonian
is proportional to the square of the supercharge (from supersymmetry), 
one may
formulate the  bound state problem in terms of
the supercharge \cite{mss95,hak95}.

A particular motivation for studying $1+1$ dimensional 
field theories in the light-cone
formalism is the simplicity of certain bound states -- the t'Hooft pion
and Schwinger particle being well known examples of this. Analogs of the 
t'Hooft pion in a non-supersymmetric theory
involving (complex) adjoint fermions have also been discovered
\cite{anp97,pin97}. All of these bound states are characterized
by relatively simple Fock state expansions, and in particular, 
there is an upper bound on the
allowed number of partons appearing in each Fock state. 
It is therefore
of interest to see whether the
massless states in the DLCQ  \cite{dlcqpapers} formulation of the model
studied here also
admit simple Fock state expansions
in the continuum limit.
We will be addressing that question in detail here, while
providing a more thorough discussion of our numerical results 
elsewhere \cite{alp98a}.

We should stress that a number of
recent string theory developments have sharpened the need to understand
supersymmetric
Yang-Mills in various dimensions, since they play
a crucial role in describing D-brane dynamics 
(see \cite{tay98} for a review),
and, ultimately, in formulating M(atrix) Theory \cite{bfss97}. 
An interpretation
of the matrix model for M(atrix) Theory at finite $N$ has also been
given by Susskind \cite{suss97}, providing additional motivation
to study super Yang-Mills at finite $N$.
We should stress, however, that in the model we study here, 
we compactify the null direction $x^-$, rather 
than in a spatial
direction.    
Furthermore, we drop the zero mode sector \cite{pin97a,mrp97}, 
which is conventional in DLCQ, 
and therefore we eliminate any possibility of
connecting our solutions with an equal time quantization
of the same theory with
a spatially compactified dimension.
However, experience with DLCQ has shown that the massive
spectrum is
insensitive to how the theory is compactified, and to the zero modes.  For
the massless
spectrum this may not be the case.

The organization of the paper proceeds as follows; in 
Section \ref{formulation}
we formulate the bound state problem 
for a two dimensional matrix model 
in the light-cone formalism. In particular, we write down
explicit expressions for the quantized supercharges of this theory.
In Section \ref{proof}, we present an analytical study of the continuum 
bound state equations for the gauge group SU($N$), 
and conclude that there can be no normalizable SU($N$)
bound state with an upper limit on the number of partons in
its Fock state expansion. Remarkably, the proof hinges 
on the assumption that the eigenstates are normalizable - no further
properties concerning the eigenstate wave functions are needed in addition
to the fact that they satisfy the bound state equations.
The proof is given in several steps. First we consider the 
validity of this proposition in the large $N$ approximation, 
and for massless bound states. We then 
generalize the proof for finite $N$, and for massive states. 
In Section \ref{dlcq} we
discuss in detail the massless solutions that appear in the 
DLCQ bound state equations.
 Finally, in Section \ref{conclusions},
we review some of the
implications of our results with regards to the utility
of DLCQ as a non-perturbative approach towards solving Yang-Mills
field theories at finite and large $N$.

\section{ \bf Formulation of the bound state problem.}
\label{formulation}
\renewcommand{\theequation}{2.\arabic{equation}}
\setcounter{equation}{0}

The light-cone formulation of the 
supersymmetric matrix model obtained by dimensionally 
reducing ${\cal N} = 1$ $\mbox{SYM}_{2+1}$ to $1+1$ dimensions
has already appeared in \cite{mss95}, to which we refer the reader for
explicit derivations. We simply note here that
the light-cone Hamiltonian $P^-$ is given in terms of the 
supercharge $Q^-$ via the supersymmetry
relation $\{Q^-,Q^-\} = 2 \sqrt{2} P^-$, where
\begin{equation}
    Q^-  =  2^{3/4} g \int dx^- \mbox{tr} \left\{
         ({\rm i}[\phi,\partial_- \phi ] + 2 \psi \psi ) \frac{1}{
    \partial_-} \psi \right\}. \label{qminus}
\end{equation}
In the above, $\phi_{ij} = \phi_{ij}(x^+,x^-)$ and 
$\psi_{ij} = \psi_{ij}(x^+,x^-)$
are $N \times N$ Hermitian matrix fields representing the physical
boson and fermion degrees of freedom (respectively) of the theory,
and are remnants of the physical transverse degrees of freedom
of the original $2+1$ dimensional theory. 
This is a special feature of light-cone quantization in light-cone 
gauge: all unphysical degrees of freedom present in the original
Lagrangian may be explicitly eliminated. There are no ghosts.

For completeness, we indicate the additional relation 
$\{Q^+,Q^+\} = 2 \sqrt{2} P^+$ for the light-cone momentum $P^+$,
where 
\begin{equation}
    Q^+  =  2^{1/4} \int dx^- \mbox{tr} \left[
         (\partial_- \phi)^2 + {\rm i}  \psi \partial_- \psi   \right].
\end{equation}
The $(1,1)$ supersymmetry of the model follows from the fact 
 $\{Q^+,Q^-\} = 0$.
%
%
%
%
%
%
In order to quantize $\phi$ and $\psi$ on the light-cone, we
first introduce the following
expansions at fixed light-cone time $x^{+}=0$:
\begin{eqnarray}
\phi_{ij}(x^-,0)=\frac{1}{\sqrt{2\pi}}\int_0^{\infty}
\frac{dk^+}{\sqrt{2k^+}}\left(a_{ij}(k^+) e^{-ik^+ x^-}+a^\dagger_{ji}(k^+)
e^{ik^+ x^-}\right); \label{phiexp}\\
\psi_{ij}(x^-,0)=\frac{1}{2\sqrt{\pi}}\int_0^{\infty}
dk^+ \left(b_{ij}(k^+) e^{-ik^+ x^-}+b^\dagger_{ji}(k^+)
e^{ik^+ x^-}\right). \label{psiexp}
\end{eqnarray}
We then specify the commutation relations
\begin{equation}
\left[a_{ij}(p^+ ),a^\dagger_{lk}(q^+)\right]=\left\{ b_{ij}(p^+ ),
b^\dagger_{lk}(q^+)\right\}=\delta (p^+ -q^+)\del_{il}\del_{jk} 
\end{equation}
for the gauge group U($N$), or
\begin{equation}
\left[a_{ij}(p^+ ),a^\dagger_{lk}(q^+)\right]=\left\{ b_{ij}(p^+ ),
b^\dagger_{lk}(q^+)\right\}=\delta (p^+ -q^+)\left(\del_{il}\del_{jk}-
\frac{1}{N}\del_{ij}\del_{kl}\right)
\end{equation}
for the gauge group SU($N$)\footnote{We assume the normalization 
${\mbox tr}[T^a T^b] = \delta^{ab}$, where the $T^a$'s are the 
generators of the Lie algebra of SU($N$).}.  

For the bound state eigen-problem
$2P^+ P^- |\Psi> = M^2 |\Psi>$, we may restrict to the
subspace of states with fixed light-cone momentum $P^+$, 
on which $P^+$ is diagonal, and so the bound state problem is 
reduced to the diagonalization
of the light-cone Hamiltonian $P^-$.
Since $P^-$ is proportional to the square of the supercharge $Q^-$,
any eigenstate $|\Psi>$ of $P^-$ with mass squared $M^2$ gives
rise to a natural four-fold degeneracy in the spectrum because
of the supersymmetry algebra---all four states below have the same mass: 
\begin{equation}
    |\Psi>, \hspace{4mm} Q^+ |\Psi>,\hspace{4mm}  Q^- |\Psi>, 
\hspace{4mm}  Q^+ Q^- |\Psi>.  
\end{equation}
Although this four-fold degeneracy is realized in the continuum
formulation of the theory, this property will not
necessarily survive if we choose to discretize the 
theory in an arbitrary manner. However, a nice  
feature of DLCQ is that it does
preserve the supersymmetry (and hence the {\it exact} four-fold
degeneracy) for any resolution.  

Focusing attention on zero mass eigenstates, we simply note
that a massless eigenstate of $P^-$ must also be annihilated by the 
supercharge $Q^-$, since $P^-$ is proportional to $(Q^-)^2$. Thus
the relevant eigen-equation is $Q^- |\Psi> = 0$. We wish to study
this equation. 
However, first we need to state the explicit equation for $Q^-$,
in the momentum representation, 
which is obtained by substituting 
the quantized field expressions (\ref{phiexp}) and (\ref{psiexp})
directly into the the definition of the supercharge (\ref{qminus}).
The result is:
\begin{eqnarray}
\label{Qminus}
Q^-&=& {{\rm i} 2^{-1/4} g \over \sqrt{\pi}}\int_0^\infty dk_1dk_2dk_3
\delta(k_1+k_2-k_3) \left\{ \frac{}{} \right.\nonumber\\
&&{1 \over 2\sqrt{k_1 k_2}} {k_2-k_1 \over k_3}
[a_{ik}^\dagger(k_1) a_{kj}^\dagger(k_2) b_{ij}(k_3)
-b_{ij}^\dagger(k_3)a_{ik}(k_1) a_{kj}(k_2) ]\nonumber\\
&&{1 \over 2\sqrt{k_1 k_3}} {k_1+k_3 \over k_2}
[a_{ik}^\dagger(k_3) a_{kj}(k_1) b_{ij}(k_2)
-a_{ik}^\dagger(k_1) b_{kj}^\dagger(k_2)a_{ij}(k_3) ]\nonumber\\
&&{1 \over 2\sqrt{k_2 k_3}} {k_2+k_3 \over k_1}
[b_{ik}^\dagger(k_1) a_{kj}^\dagger(k_2) a_{ij}(k_3)
-a_{ij}^\dagger(k_3)b_{ik}(k_1) a_{kj}(k_2) ]\nonumber\\
&& ({ 1\over k_1}+{1 \over k_2}-{1\over k_3})
[b_{ik}^\dagger(k_1) b_{kj}^\dagger(k_2) b_{ij}(k_3)
+b_{ij}^\dagger(k_3) b_{ik}(k_1) b_{kj}(k_2)]  \left. \frac{}{}\right\}.
\end{eqnarray}

In order to implement the DLCQ formulation \cite{dlcqpapers}
of the theory, we simply restrict the 
momenta $k_1,k_2$ and $k_3$ appearing in the above equation to
the following set of allowed momenta: $\{\frac{P^+}{K},\frac{2P^+}{K},
\frac{3P^+}{K},\dots \}$. Here, $K$ is some arbitrary positive integer,
and must be sent to infinity if we wish to recover the continuum 
formulation of the theory. The integer $K$ 
is called the {\em harmonic resolution},
and $1/K$ measures the coarseness of our discretization\footnote{Recently,
Susskind has proposed a surprising connection between the 
harmonic resolution arising from the DLCQ of $M$ theory, and the
integer $N$ appearing in the U($N$) gauge group for M(atrix) Theory
(namely, they are the same) \cite{suss97}.}.
Physically, $1/K$ represents the smallest unit of longitudinal
momentum fraction allowed for each parton. 
As soon as we implement the DLCQ procedure, which is 
specified unambiguously
by the harmonic resolution $K$, the integrals appearing
in the definition of $Q^-$ are replaced by finite sums,
and the eigen-equation $Q^- |\Psi> = 0$ is reduced to a finite matrix
problem. For sufficiently small values of $K$ (in this case 
for $K \leq 4$)
this eigen-problem may be solved analytically.
For values $K > 5$, we may still compute
the DLCQ supercharge analytically as a function of $N$,
but the diagonalization procedure must be performed 
numerically. A detailed discussion of the DLCQ 
analytical and numerical results 
of this work will appear elsewhere \cite{alp98a}.

For now, we concentrate on the structure of the zero mass eigenstates 
for the continuum theory. Firstly, note 
that for the U($N$) bound state problem, massless states
appear automatically because of the decoupling of the U($1$)
and SU($N$) degrees of freedom that constitute U($N$). More explicitly,
we may introduce the U(1) operators
\begin{equation}
         \alpha (k^+)  =  \frac{1}{N}\mbox{tr} [a(k^+)] 
\hspace{4mm} \mbox{and} 
   \hspace{4mm} \beta (k^+)  =  \frac{1}{N}\mbox{tr} [b(k^+)],
\end{equation}
which allow us to decompose any U($N$) operator into a sum of
U(1) and SU($N$) operators:
\begin{equation}
            a(k^+) = \alpha (k^+)\cdot \mbox{${\bf 1}_{N\times N}$} + 
{\tilde a}(k^+)
 \hspace{4mm} \mbox{and} \hspace{4mm}
      b(k^+) = \beta (k^+)\cdot \mbox{${\bf 1}_{N\times N}$}
 + {\tilde b}(k^+), 
\end{equation}
where ${\tilde a}(k^+)$ and ${\tilde b}(k^+)$ are traceless $N \times N$
matrices. If we now substitute the operators above into the
expression for the supercharge (\ref{Qminus}), 
we find that
all terms involving the U(1) factors $\alpha(k^+), \beta(k^+)$ vanish -- only
the SU($N$) operators ${\tilde a}(k^+),{\tilde b}(k^+)$ survive.
i.e. starting with the definition of the U($N$)
supercharge, we end up with the definition of the SU($N$) supercharge.
In addition, the (anti)commutation relations
$[{\tilde a}_{ij}(k_1),\alpha^{\dagger}(k_2)] = 0$ and
$\{{\tilde b}_{ij}(k_1),\beta^{\dagger}(k_2)\} = 0$ imply 
that this supercharge acts only on the SU($N$) creation operators
of a fock state - the U(1) creation operators only introduce
degeneracies in the SU($N$) spectrum. Clearly, since $Q^-$ has no
U(1) contribution, any fock state made up of only U(1)
creation operators must have zero mass.
 The non-trivial problem here is to determine whether there are 
massless states for the SU($N$) sector. We will address 
this topic next.

\section{ \bf The Proof}
\label{proof}
\renewcommand{\theequation}{3.\arabic{equation}}
\setcounter{equation}{0}

It was pointed out in the previous section
that a zero mass eigenstate is annihilated by the 
light-cone supercharge (\ref{Qminus}):
\begin{equation}
\label{mslcond}
Q^-|\Psi \rangle =0
\end{equation}
We wish to show that if such an SU($N$) 
eigenstate is normalizable, then
it must involve a superposition of an {\it infinite} number of Fock states.
The basic strategy is quite simple; normalizability will 
impose certain conditions on the light-cone wave functions as one or
several momentum variables vanish. Moreover, if we assume a given
eigenstate $|\Psi \rangle$ has at most $n$ partons, then the terms in 
$Q^- |\Psi \rangle$ consisting of $n+1$ partons must sum to zero,
providing relations between the $n$ parton wave functions only.
We then show these wave functions must all vanish by studying 
various zero momentum limits of these relations. Interestingly, the 
utility of studying light-cone wave functions at small momenta
also appears in the context of light-front 
$\mbox{QCD}_{3+1}$ \cite{adb97}. 

In order to proceed with a systematic presentation of
the proof, we start by considering the large $N$ limit case.
This simply means that we consider Fock states that 
are made from a {\em single} trace of a product 
of boson or fermion creation
operators acting on the light-cone Fock vacuum
$|0\rangle$. Multiple trace states correspond to $1/N$ corrections to 
the theory, and are therefore ignored.
In this limit, a general state $|\Psi \rangle$ is
a superposition of Fock states of any length, 
and may be written in the form   
\begin{eqnarray}
\label{genfock}
\lefteqn{|\Psi \rangle =
  \sum_{n=2}^{\infty} \sum_{r=0}^{n}
   \sum_{P} \int_0^{P^+} {dq_1 \dots dq_n \over 
   \sqrt{q_1...q_n}}\delta(q_1 + \cdots + q_n - P^+) \times} & &  
\nonumber \\
& & \hspace{25mm} 
f^{(n,r)}_P (q_1,\dots,q_n)
 \mbox{tr}[c^\dagger(q_1)\dots c^\dagger(q_n)]|0\rangle ,
\end{eqnarray}
where $c^{\dagger}(q^+)$ represents either a boson or fermion 
creation operator
carrying light-cone momentum $q^+$, and $f^{(n,r)}_P$ denotes 
the wave function of an $n$ parton Fock state containing $r$
fermions in a particular arrangement $P$. It is implied that
we sum over all such arrangements, which may not necessarily be distinct
with respect to cyclic symmetry of the trace.  

At this point, we simply remark that normalizability of a
general state $|\Psi \rangle$ above implies
\begin{equation}
\int_0^{P^+} {dq_1 \dots dq_n \over q_1 \dots q_n}\delta(
q_1 + \cdots + q_n - P^+)|f^{(n,r)}_P(q_1,\dots,q_n )|^2 < \infty
\end{equation}
for any particular wave function $f^{(n,r)}_P$. Therefore,
any wave function vanishes 
if one or several of its momenta are made to
vanish.

We are now ready to carry out the details of the proof. But first 
a little notation. We will write $|\Psi_{(n,m)} \rangle$ to denote 
a superposition of all Fock states -- as in (\ref{genfock}) --
with precisely $n$ partons, $m$ of which are fermions. Such a 
Fock expansion involves only the wave functions $f^{(n,m)}_P$, 
and the number of them is
enumerated by the index $P$. 
For the special
case $|\Psi_{(n,0)} \rangle$ (i.e. no fermions), there 
is only one wave function, which we denote by $f^{(n,0)}$
for brevity:
\begin{equation}
\label{stallbos}
|\Psi_{(n,0)} \rangle =
  \int_0^{P^+} {dq_1 \dots dq_n \over 
   \sqrt{q_1...q_n}}\delta(q_1 + \cdots + q_n - P^+)\hspace{1mm}  
f^{(n,0)}(q_1,\dots,q_n)
 \mbox{tr}[a^\dagger(q_1)\dots a^\dagger(q_n)]|0\rangle .
\end{equation}
There is another special case we wish to consider; namely, 
the state $|\Psi_{(n,2)} \rangle$ consisting of $n$ parton
Fock states with precisely two fermions. If we place one of
the fermions at the beginning of the trace, then there are 
$n-1$ ways of positioning the second fermion, yielding
$n-1$ possible wave functions. We will enumerate such
wave functions by the subscript index $k$, as in $f^{(n,2)}_k$,
where $k=2,3,\dots,n$.
The subscript $k$ denotes the location of the second fermion.
Explicitly, we have
\begin{eqnarray}
\label{twofermion}
\lefteqn{
|\Psi_{(n,2)} \rangle =
 \sum_{k=2}^{n} \int_0^{P^+} {dq_1 \dots dq_n \over 
   \sqrt{q_1...q_n}}\delta(q_1 + \cdots + q_n - P^+)\times } & &\nonumber \\
& & \hspace{15mm}  
f^{(n,2)}_k(q_1,\dots,q_k,\dots,q_n)
 \mbox{tr}[ b^\dagger(q_1)a^\dagger(q_2)\dots
   b^\dagger(q_k) \dots
a^\dagger(q_{n})]|0\rangle .
\end{eqnarray}
Of course, depending upon the symmetry, the $n-1$
Fock states enumerated in this way need not be distinct 
with respect to the cyclic properties of the trace.
This provides us with additional relations between wave functions -- a fact
we will make use of later on.
  
Now let us assume that $| \Psi \rangle$ is a
normalizable SU($N$) zero mass eigenstate with at most $n$ partons.
Glancing at the form of (\ref{Qminus}), we see that
the $n+1$ parton Fock states containing a single
fermion in each of the combinations
 $Q^-|\Psi_{(n,0)} \rangle$ and  $Q^-|\Psi_{(n,2)} \rangle$
must cancel each other to guarantee a massless eigenstate.
This immediately
gives rise to the following wave function relation:
\begin{eqnarray}
\label{higheqnnorm}
\frac{q_{1}+2q_2}{q_{1}+q_2} f^{(n,0)}(q_1+q_2,q_3,\dots,q_{n+1})
-\frac{q_1+2q_{n+1}}{q_1+q_{n+1}}f^{(n,0)}(q_{n+1}+q_1,q_2,\dots,q_n)
=\nonumber\\
=2 \frac{\sqrt{q_1}}{n}\sum_{k=2}^{n} \frac{q_{k+1}-q_k}{(q_{k+1}+q_k)^{3/2}}
f^{(n,2)}_k(q_1,\dots,q_{k-1},q_k+q_{k+1},q_{k+2},\dots,q_{n+1}).
\end{eqnarray}
In the limit $q_i \rightarrow 0$, for $3 \leq i \leq n$, this last equation
is reduced to  
\begin{eqnarray}
\lefteqn{
\frac{1}{\sqrt{q_{i+1}}}f^{(n,2)}_i(q_1,\dots,q_{i-1},q_{i+1},
\dots,q_{n+1}) } & & \nonumber \\
& & \hspace{25mm} -
\frac{1}{\sqrt{q_{i-1}}}f^{(n,2)}_{i-1}(q_1,\dots,q_{i-1},q_{i+1},
\dots,q_{n+1})=0.
\end{eqnarray}
An immediate consequence is that
any wave function $f^{(n,2)}_i$ for $i=3,4,\dots,n$, may
be expressed in terms of $f^{(n,2)}_2$. Explicitly, we have 
\begin{equation}
f^{(n,2)}_i(q_1,q_2,\dots,q_n) = \sqrt{\frac{q_i}{q_2}}
    f^{(n,2)}_2(q_1,q_2,\dots,q_n), \hspace{8mm}i=3,4,\dots,n.
\label{chain}
\end{equation} 
Moreover, the limit $q_2 \rightarrow 0$  
of equation (\ref{higheqnnorm})
yields the further relation after a suitable change of variables:
\begin{eqnarray}
f^{(n,0)}(q_1,q_2,q_3,\dots,q_{n}) & = &\frac{2}{n}\sqrt{\frac{q_{1}}{q_{2
}}}
f^{(n,2)}_2(q_1,q_2,q_3,\dots,q_{n}). \label{rel1}
\end{eqnarray}
Finally, because of the cyclic properties of the trace, there 
is an additional relation between wave functions:
\begin{equation}
  f^{(n,2)}_i(q_1,q_2,\dots,q_i,\dots,q_n)
 = -f^{(n,2)}_{n-i+2}(q_i,q_{i+1},\dots,q_n,q_1,q_2,\dots,q_{i-1}).
\end{equation}
Setting $i=2$ in the above equation, and $i=n$ in equation (\ref{chain}),
we deduce 
\begin{equation} 
       f^{(n,2)}_2(q_1,q_2,\dots,q_n) = 
        -\sqrt{\frac{q_1}{q_2}} f^{(n,2)}_2(q_2,q_3,\dots,q_n,q_1).
\end{equation}
Combining this with equation (\ref{rel1}), we conclude
$(\frac{\sqrt{q_2}}{q_1} + \frac{\sqrt{q_3}}{q_2})
f^{(n,0)}(q_1,\dots,q_n) = 0$, where
we use the fact that the wave functions $f^{(n,0)}$ are 
cyclically symmetric.  Thus $f^{(n,0)}$ must vanish.  
It immediately follows that $f^{(n,2)}_i$ vanish for all $i$
as well.  

To summarize, we have shown that if $|\Psi \rangle$ is a normalizable 
zero mass eigenstate, where each Fock state
in its Fock state expansion has no more than $n$ partons,
the contributions 
$|\Psi_{(n,0)}\rangle$ and $|\Psi_{(n,2)}\rangle$ 
in this Fock state expansion must vanish. Since we may
assume $|\Psi \rangle$ is bosonic, the only other
contributions involve Fock states with an even number of
fermions: $|\Psi_{(n,4)}\rangle$, $|\Psi_{(n,6)}\rangle$, and so on.
We claim that all such contributions vanish. To
see this, first observe that the $n+1$ parton Fock states
with three fermions
in the combinations $Q^- |\Psi_{(n,2)}\rangle$ and 
$Q^- |\Psi_{(n,4)}\rangle$ must cancel each other, in order
to guarantee a zero eigenstate mass. But our previous analysis
demonstrated that $|\Psi_{(n,2)}\rangle  \equiv 0$, and thus
the $n+1$ parton Fock states with three fermions in
$Q^- |\Psi_{(n,4)}\rangle$ alone must sum to zero.

We are now ready to perform an
induction procedure. Namely,  
we assume  that for some positive integer
$k$ the state $|\Psi_{(n,2[k-1])}\rangle$ vanishes. 
Then the $n+1$ parton Fock states
in $Q^- |\Psi \rangle$ which contain $2k-1$ fermions 
receive contributions only from  $Q^-|\Psi_{(n,2k)}\rangle$
in which a fermion is replaced by two bosons. This has to sum to zero.
We therefore obtain a relation among the wave functions
$f^{(n,2k)}_P$ by considering the action
of the supercharge (\ref{Qminus}) in which a fermion is replaced
by two bosons. Keeping in mind that we are
 free to renormalize any wave function by a constant, 
we end up with the following relation:
\begin{equation}
\label{indeqnnorm}
\sum_{P} {f}^{(n,2k)}_P(s_1,\dots,s_{i-1},s_i+s_{i+1},s_{i+2},
\dots, s_{n+1}) \frac{s_{i+1}-s_i}{(s_{i+1}+s_i)^{3/2}}=0.
\end{equation}
It is now an easy task to show that 
 the wave functions ${f}^{(n,2k)}_P$ appearing
in equation (\ref{indeqnnorm}) must vanish; one simply  
considers various limits $s_j \rightarrow 0$ as we did before.
This completes our proof by induction. Namely, there
can be no non-trivial normalizable massless state
with an upper limit on the number of allowed partons.
Of course, this proof is valid only in the large $N$ limit.
We now turn our attention to the finite $N$ case.

\medskip

Let us define $Q^-_{lead}$ to be that part of the 
supercharge $Q^-$ that replaces a fermion with two bosons,
or replaces a boson with a boson and fermion pair.
As in the large $N$ case we begin
by assuming that we have a normalizable zero mass eigenstate 
$|\Psi \rangle$ which is a sum of Fock states that have 
at most $n$ partons.
The proof for finite $N$ consists of
two parts. First, we consider bosonic states 
consisting of only $n$ parton Fock states
that have at most two fermions,
and show the wave functions must
vanish. We then invoke an induction argument to 
consider $n$ parton wave functions involving an even number of fermions,
and show they must vanish as well.

The additional complication introduced by the assumption that $N$ is 
finite is that a given Fock state may involve more than just
a single trace. However, note that
$Q^-_{lead}$
cannot decrease the number of traces; it can either increase the
number of traces by one, or leave the number unchanged. 
Thus we have a natural induction
procedure in the number of traces as well.
Since the terms in $Q^-_{lead}$  have
 only one annihilation operator, it acts on
a given product of traces according to the Leibniz rule:
\begin{equation}
Q^-_{lead}\left(\mbox{tr}[A]\mbox{tr}[B]\dots \right)|0\rangle =
\left(Q^-_{lead}\mbox{tr}[A]\right)\mbox{tr}[B]\dots|0>+
(-1)^{F(A)}\mbox{tr}[A]Q^-_{lead}\left(\mbox{tr}[B]\dots\right)|0>.
\end{equation}
Schematically, the general structure of an arbitrary Fock
state with $k$ traces has the form
\begin{equation}
f^{(n,i_1,i_2,\dots,i_k)}_P 
\mbox{tr}[(b^\dagger)^{i_1} a^\dagger\dots a^\dagger]
\dots \mbox{tr}[(b^\dagger)^{i_k} a^\dagger \dots a^\dagger]|0>,
\end{equation}
where  $n$ denotes 
the total number of partons in each Fock state, 
and the integers $i_1,i_2,\dots$
denote the number of fermions in the first trace, second trace, and so on.
We will always
order the traces so that the number of fermions
in each trace  
decreases to the right. The index $P$ labels a particular arrangement
of fermions.

We now consider the $n+1$ parton Fock states of
$Q^-_{lead}|\Psi \rangle$ that have precisely one fermion.
The only possible contributions involve three types of wave functions;
$f^{(n,0)}$, $f^{(n,2)}_P$ and
$f^{(n,1,1)}$ (we only include the permutation index $P$
if there is more than one distinct arrangement).
If these three wave functions  contribute to the 
same one fermion Fock state, then the distribution of
bosons in the Fock state corresponding to
$f^{(n,2)}_P$ determines the distribution of bosons for $f^{(n,0)}$ and
$f^{(n,1,1)}$. We allow
$Q^-_{lead}$ to act only on the first trace in both
$f^{(n,0)}$ and $f^{(n,2)}_P$,
and only on the second one in $f^{(n,1,1)}$. If
there are more than two traces in these states 
they must be identical in all the
components, and so don't play a role in the calculation.
Thus, it is sufficient to
consider states with two traces only. Such a state has the form
\begin{eqnarray}
\label{finnphi}
|\Phi> &=&\int_0^{P^+} { d^{m+n}q \over \sqrt{q_1 \dots q_{n+m}}} 
\delta(q_1+ \cdots +q_{n+m}
-P^+)
\nonumber \\
& &f^{(n+m,0)}(q_1,\dots,q_m|q_{m+1},\dots,q_{m+n})
\nonumber \\
&\times &\mbox{tr}[a^\dagger(q_1)\dots a^\dagger(q_m)]
\mbox{tr}[a^\dagger(q_{m+1})\dots a^\dagger(q_{m+n})]|0\rangle
\nonumber\\
&+&\int_0^{P^+} { d^{m+n-2}qdp_1dp_2 \over \sqrt{q_1\dots q_{n+m-2}p_1p_2}}
\delta(q_1+\cdots +q_{n+m-2}+p_1+p_2 -P^+) \{
\nonumber \\
& &f^{(n+m,1,1)}(p_1,q_1,\dots,q_m|p_2,q_{m+3},\dots,q_{m+n}) \times
\nonumber\\
&\times &\mbox{tr}[b^\dagger(p_1)a^\dagger(q_1)\dots a^\dagger(q_m)]
\mbox{tr}[b^\dagger(p_2)a^\dagger(q_{m+3})\dots a^\dagger(q_{m+n})]
 +
\nonumber\\
&+&\sum_P
f^{(n+m,2)}_P (p_1,P[q_1\dots q_{m-2};p_2]|q_{m+1}\dots q_{m+n}) \times
\nonumber \\
&\times &\mbox{tr}\left(b^\dagger(p_1)
P[a^\dagger(q_1)\dots a^\dagger(q_{m-2});b^\dagger(p_2)]
\right)
\mbox{tr}[a^\dagger(q_{m+1})\dots a^\dagger(q_{m+n})] \}|0\rangle,
\end{eqnarray}
\\
\noindent where we have  summed over the index
$P$ representing all possible permutation arrangements
 of bosons and fermions
that
contribute.
We then
find,
\begin{eqnarray}
\lefteqn{ {F}(p,q_1,\dots,q_m|q_{m+1},q_{m+2},\dots,q_{m+n})+ } 
& &\nonumber\\
& & +\frac{q_{m+2}-q_{m+1}}{(q_{m+2}+q_{m+1})^{3/2}}
{f}^{(n+m,1,1)}(p,q_1,\dots,q_m|q_{m+1}+q_{m+2},q_{m+3},\dots,q_{m+n})=0,
\end{eqnarray}
where $F$ is the contribution from $f^{(n+m,0)}$ and $f^{(n+m,2)}_P$.
Now we see that the limit $q_{m+1}\rightarrow 0$ gives:
$f^{(n+m,1,1)}\equiv 0$.
Thus if (\ref{finnphi}) represents 
a contribution to the massless eigenstate state $|\Psi\rangle$,
then $|\Phi\rangle$ takes the form
\begin{eqnarray}
\label{finnphm}
|\Phi\rangle &=& \int_0^{P^+} {d^{m+n-2}q dK^+ 
 \over \sqrt{q_1\dots q_{n+m-2}}}
\delta(q_1+ \cdots + q_{n+m-2} - (P^+-K^+))\left[ \frac{}{} \right.
\nonumber \\
&&  \int_0^{P^+}
{dq_{m-1}dq_{m} \over \sqrt{q_{m-1}q_m}}
\delta(q_{m-1} +q_m -K^+)
\nonumber \\
& & f^{(n+m,0)}(q_1,\dots,q_m|q_{m+1},\dots,q_{m+n})
\mbox{tr}(a^\dagger(q_1)\dots a^\dagger(q_m))
\nonumber\\
&+&  \int_0^{P^+} {dp_1dp_1 \over \sqrt{p_1p_2}} \delta(p_1+p_2 -K^+)
\nonumber \\
&& \sum_P f^{(n+m,2)}_P(p_1,P[q_1,\dots,q_{m-2};p_2]|q_{m+1},\dots,q_{m+n})
\nonumber \\
&&\left.
\mbox{tr}(b^\dagger(p_1)P[a^\dagger(q_1)\dots
a^\dagger(q_{m-2});b^\dagger(p_2)])
\mbox{tr}(a^\dagger(q_{m+1})\dots a^\dagger(q_{m+n}))\right]|0\rangle
\end{eqnarray}
and $Q^-_{lead}$ acts only on the terms in the square brackets. All these
terms have only one trace, which is a scenario we already
encountered  in
the large $N$ limit case. 
Using the results of that discussion, we find that the only
massless solution of the form (\ref{finnphm}) is the trivial one. This is the
starting point of the induction procedure for finite $N$.

As explained earlier, we look for $n$ parton
Fock states in the expansion for $|\Psi\rangle$ that have $2k$ fermions
($k>1$),
 To finish the proof we
need to
show that for any $k$ the only allowed wave function is the trival one.
>From the
large $N$ result we know there are no such one trace states. We now
consider the state
with an arbitrary number of traces,
\begin{eqnarray}
|\Psi_{(n,2k)}>&=&\sum_{P} \int_0^{P^+} {ds_1\dots ds_n \over \sqrt{s_1\dots s_n}}
\delta  (s_1+\cdots + s_n -P^+) \nonumber \\ &&f^{(n,2k)}_P
(s_1\dots s_{i_1}|\dots |\dots s_n)
\mbox{tr}\left(c^\dagger(s_1)\dots c^\dagger(s_{i_1})\right) 
\mbox{tr}\left(\dots \right)
\mbox{tr}\left(\dots c^\dagger(s_n)\right)|0>,
\end{eqnarray}
then the analog of (\ref{indeqnnorm}) for such states reads:
\begin{equation}
\label{finnind}
{\sum_{i}}'  f^{(n,2k)}_{P_i}
(s_1\dots |s_{j_a} \dots s_{i-1},s_i+s_{i+1},s_{i+2}\dots
s_{j_a+k_a}| \dots s_{n+1})
\frac{s_{i+1}-s_i}{(s_{i+1}+s_i)^{3/2}}=0.
\end{equation}
Here, $\sum_{i}'$ means that for each trace we should include one additional
term with $"i"=j_a+k_a$, $"i+1"=j_a$ if $c$ corresponding to both $j_a+k_a$
and $j_a$ is  $a$. If the number of traces is $a$, we introduce
$$
j_a =\sum_{b=1}^{a-1} k_b.
$$

If any of the blocks $\mbox{tr}(\dots)$ in the state for which 
(\ref{finnind})
is written contains two or more fermions, then, as in the large $N$ case,
all the corresponding wave functions 
$f^{(n,2k)}_{P}$ vanish. So we only need to consider the states
of the form:
\begin{eqnarray}
\label{st11b}
\lefteqn{|\Psi_{(n,k_1+1,\dots)}\rangle
= \sum_P\int dpdq f^{(n,k_1+1,\dots)}_P
(p_1,q_1,\dots,q_{k_1}|p_2,q_{k_1+1},\dots,
q_{k_1+k_2}|\dots)  \times} & & \nonumber \\
& & \mbox{tr}\left(b^\dagger(p_1)a^\dagger(q_1)\dots a^\dagger(q_{k_1})\right)
\mbox{tr}\left(b^\dagger(p_2)a^\dagger(q_{k_1+1})\dots
a^\dagger(q_{k_1+k_2})
\right)\dots |0\rangle.
\end{eqnarray}
Let ${\tilde Q}$ denote that part of the supercharge $Q^-$
which replaces a fermion with two bosons.
 Let us consider the result of such a change in the
first trace. Suppose there are $a$ traces having the same form as the
first trace. Then without loss of generality, we may assume
they are the first
$a$ traces. Then using the symmetries of the wave functions we find:
\begin{eqnarray}
&&{\tilde Q}|\Psi_{(n,k_1+1,\dots)}\rangle  =-\frac{1}{2\sqrt{2\pi}}
\sum_P
\int_0^{P^+} dkdpdq f^{(n,k_1+1,\dots)}_P
(p_1,q_1,\dots,q_{k_1}|p_2,q_{k_1+1},\dots,q_{2k_1}|\dots) \times \nonumber \\
&& \hspace{25mm}
\sum_{b=1}^{a} \frac{p_b-2k}{p_b}\frac{1}{\sqrt{k(p_b-k)}}(-1)^{b+1}
\times \nonumber\\
&&\mbox{tr}
\left(b^\dagger(p_1)a^\dagger(q_1)\dots a^\dagger(q_{k_1})\right)\dots
\mbox{tr}\left(a^\dagger(k)a^\dagger(p_b-k)a^\dagger(q_{(b-1)k_1+1})\dots
a^\dagger(q_{bk_1})\right)
\dots |0\rangle \nonumber\\
&&=-\frac{1}{2\sqrt{2\pi}}
\sum_P\int_0^{P^+} dkdpdq \frac{p_1-2k}{p_1}\frac{1}{\sqrt{k(p_1-k)}}
\mbox{tr}\left(a^\dagger(k)a^\dagger(p_1-k)a^\dagger(q_1)\dots
a^\dagger(q_{k_1})\right) \times \nonumber \\
&&\mbox{tr}\left(b^\dagger(p_2)a^\dagger(q_{k_1+1})\dots
a^\dagger(q_{k_1+k_2})\right)
\dots |0>
 \sum_{b=1}^{a} (-1)^{b+1}(-1)^{b+1} \times \nonumber \\
&& \hspace{10mm}f_P^{(n,k_1+1,\dots)}
(p_1,q_1,\dots,q_{k_1}|p_2,q_{k_1+1},\dots,q_{k_1+k_2}|\dots).\nonumber
\end{eqnarray}
If the above expression vanishes then the only solution is the trivial
one in which all wave functions vanish. This
finishes the proof of the induction procedure for the finite $N$ case.

The extension of the proof to massive bound states is straightforward.
Firstly, assume $|\Psi\rangle$ is a normalizable
 eigenstate of $2P^+ P^-$
with mass squared $M^2 \neq 0$.
Then, since $P^- = \frac{1}{\sqrt{2}}(Q^-)^2$,
the state  
\begin{equation}
          |{\tilde \Psi}\rangle \equiv |\Psi\rangle
                           + \alpha Q^- |\Psi\rangle
\end{equation} 
for $\alpha^2 = \sqrt{2} P^+/M^2$ is a normalizable eigenstate 
of the supercharge $Q^-$, with eigenvalue $1/\alpha$.
We therefore study the eigen-problem $Q^-|{\tilde \Psi}\rangle = 
\frac{1}{\alpha} |{\tilde \Psi}\rangle$.
The resulting constraints on the wave functions
may be obtained by  
modifying  our original expressions by including a 
wave function multiplied by a finite constant. However,
in our analysis, we always need to
let some of the momenta vanish, and therefore
this additional contribution vanishes. The analysis (and therefore
the conclusions) remains unchanged.

We therefore conclude that any normalizable SU($N$)
bound state (massless or massive) 
that exists in the model must be a superposition
of an infinite number of Fock states.

\section{ \bf Bound States in DLCQ.}
\label{dlcq}
\renewcommand{\theequation}{5.\arabic{equation}}
\setcounter{equation}{0}

In the previous section we proved that the continuum
formulation of the theory does not have any
normalizable bound states with a finite number of partons. 
Our proof used the
behavior of wave functions  at small momenta 
arising from the normalizability assumption. Neither
of these
properties can be used in DLCQ, however. Here we consider some simple
examples of
massless DLCQ solutions with $n$ bosons to help shed some light on the
relation between
DLCQ solutions and the solutions of the continuum theory.
For simplicity, we work in the large $N$ limit case.

We write the
momentum of a state in DLCQ in terms of the momentum fraction $q_i$ where
$q_i=\frac{r_i}{r}P^+$, and the $r_i$  are positive integers. 
The wave function of
such a
state is $f^{(n,0)}(r_1,\dots,r_n)$. There  are two conditions that must be
satisfied
to show that it is massless. One is the that the coefficient of
the term with one additional fermion that is producted by the action of $Q^-$
is zero. This condition gives the relation,
\begin{equation}
\label{dscreqnorm}
\frac{2r_n+t}{r_n+t}f^{(n,0)}(r_1,\dots,r_{n-1},r_n+t)-
\frac{2r_{n-1}+t}{r_{n-1}+t} f^{(n,0)}(r_1,\dots,r_{n-1}+t,r_n)=0.
\end{equation}
where $t$ correspond to the momentum fraction of the one fermion. The second is
that the coefficient of the state with two fewer bosons and one additional
fermion which
is also produced by the action of $Q^-$ is zero. This condition
gives the relation,
\begin{equation}
\label{dcreqnxt}
\sum_{k,t}\frac{t-2k}{k(t-k)}f^{(n,0)}(r_1,\dots,r_{n-2} ,k,t-k)
\delta_{(r_{n-1}+r_{n},t)}=0.
\end{equation}

For the case where all $r_i =1$, and the total harmonic
resolution is $n$, it is
trivial that
eqn(\ref{dscreqnorm})  is satisfied since there is not enough resolution to
increase
the number of particles in the state. It is also easy to see from eqn
(\ref{dcreqnxt}) since the coefficient of the one term in the sum is zero.
Thus the
wave function $f^{(n,0)}(1,1,....1)$ is a massless state for every resolution

To discuss additional solutions it is useful to start by considering
eqn(\ref{dscreqnorm}). The case $t=1$, gives the equation
\begin{equation}
\label{discrcase1}
f^{(n,0)}(r_1,\dots,r_{n-2},r_{n-1},r_n+1)=\frac{2r_{n-1}+1}{2r_n+1}
\frac{r_n+1}{r_{n-1}+1} f^{(n,0)}(r_1,\dots,r_{n-2},r_{n-1}+1,r_n).
\end{equation}
This equation is trivial to satisfy if $r_i =1$ for all $i$. 
The contributions in
eqn(\ref{dcreqnxt}) come from the two terms in the sum, $k=1$,
$t=3$ and $k=2$, $t=3$. Each term has the same coefficient but of opposite
sign and cancel. Therefore the state $f^{(n,0)}(1,...1,2)$ is a massless
state for
all resolutions

The next case  $t=2$ in eqn(\ref{dscreqnorm}) gives,
\begin{equation}
\label{discrcase2}
f^{(n,0)}(r_1,\dots,r_{n-2},r_{n-1},r_n+2)=\frac{2r_{n-1}+2}{2r_n+2}
\frac{r_n+2}{r_{n-1}+2} f^{(n,0)}(r_1,\dots,r_{n-2},r_{n-1}+2,r_n).
\end{equation}
Using (\ref{discrcase1}) twice we find:
\begin{eqnarray}
f^{(n,0)}(r_1...r_{n-2},r_{n-1},r_n+2)=\frac{2r_{n-1}+1}{2r_n+3}
\frac{r_n+2}{r_{n-1}+1} f^{(n.0)}(r_1...r_{n-2},r_{n-1}+1,
r_n+1)=\nonumber\\
=\frac{2r_{n-1}+1}{2r_n+3}\frac{r_n+2}{r_{n-1}+1}
\frac{2r_{n-1}+3}{2r_n+1}\frac{r_n+1}{r_{n-1}+2}
f^{(n,0)}(r_1...r_{n-2},r_{n-1}+2,r_n).
\end{eqnarray}
Comparing with (\ref{discrcase2}) we have:
\begin{equation}
\label{braceqn}
f^{(n,0)}(r_1...r_{n-2},r_{n-1}+2,r_n)\left(
\frac{(r_n+1)^2}{(2r_n+3)(2r_n+1)}-
\frac{(r_{n-1}+1)^2}{(2r_{n-1}+3)(2r_{n-1}+1)}\right)=0.
\end{equation}
Using relation (\ref{dscreqnorm}) several times we can always express
an arbitrary wave function in the following form:
\begin{equation}
\label{Leqn}
f^{(n,0)}(r_1...r_n)=C(r_1...r_n)f^{(n,0)}(1...1,L+1,1)
\end{equation}
where $L=r_1 +...+r_n -n$ and $C(r_1...r_n)$ is some nonzero
coefficient.The two
massless states we found above correspond to $L=0$ and $L=1$. Choosing
$r_1=...=r_{n-2}=r_n=1$ in (\ref{braceqn}) we find,
\begin{equation}
f^{(n,0)}(1...1,(L-1)+2,1)=0 \quad for \quad L>2
\end{equation}
due to monotonic behavior of the function in the parenthesis. Then using
(\ref{Leqn}) we conclude that all the wave functions with $L>2$ vanish. So the
only case we need consider is $L=2$. In this case (\ref{dscreqnorm}) has
only two nontrivial cases: $t=1$ and $t=2$ which are given by
(\ref{discrcase1}) and (\ref{discrcase2}). In the second of these equations
 we
can only have $r_1=\dots=r_n=1$ so it is trivially satisfied. Equation
(\ref{discrcase1}) however gives a nontrivial relation for the wave function:
\begin{equation}
\label{state3}
f^{(n,0)}(1,\dots,1,2,2)=f^{(n,0)}(1,\dots,2,1,2)=\dots=
f^{(n,0)}(2,\dots,1,1,2)=
\frac{10}{9}f^{(n,0)}(1,\dots,1,3).
\end{equation}
finally we must show that eqn(\ref{dcreqnxt}) is satisfied which is straight
forward.

These are only a few examples of massless states, and there are in fact many
more in
DLCQ \cite{alp98a}.
In the continuum limit we have proven that there are no massless
normalizable states with a finite number of particles.  
However, there is the possibility that at each finite value of
the harmonic resolution, one obtains an exactly massless
bound state, but as the harmonic resolution is sent to infinity, the number
of Fock states required to keep the bound state massless must
also be infinite.

\section{Conclusions}
\label{conclusions}

In this work we  
considered the dimensional reduction
of ${\cal N} = 1$ $\mbox{SYM}_{2+1}$  to
$1+1$ dimensions, and at finite $N$.
Our main objective was to analyze the structure 
of bound states both in the continuum and
in the DLCQ formulation. We discovered many 
massless states in the DLCQ formulation, but showed
that any massless (or massive) normalizable
bound states in the continuum theory must be a superposition
of an infinite number of Fock states. Our work
therefore shows
that any exact analytical treatment of the continuum
bound state problem is probably a too ambitious
objective for the near future. 
This scenario is to be 
 contrasted with
the  simple bound states discovered in a number
of $1+1$ dimensional theories with complex fermions, such as the Schwinger
model, the
t'Hooft model, and a dimensionally
reduced theory with complex adjoint fermions
\cite{anp97,pin97}. While
these theories offered hope that bound states viewed in 
the light-cone formalism might be much
simpler  than in the  equal-time quantization approach, we
 see here that this is
not the case. 
Numerical DLCQ studies of the model are, of course,
insensitive to the complexity of the bound state problem,
and extensive numerical results on the model studied here
will appear elsewhere \cite{alp98a}.

\end{sloppypar}

\begin{references}
\bibitem{bpp98} S.J. Brodsky, H.C. Pauli, and S.S. Pinsky,
{\it``Quantum Chromodynamics and Other Field Theories on the Light Cone"}
to appear in Phys.Rept. hep-ph/9705477
%
\bibitem{mss95}
Y. Matsumura, N. Sakai, and T. Sakai,
{\em Phys.Rev} {\bf D52}:2446-2461,1995 hep-th/9504150
%
\bibitem{gkm96} D. J. Gross, I. R. Klebanov, A. V. Matytsin and A. V. Smilga,
{\em Nucl.Phys} {\bf B461} (1996) 109. hep-th/9511104
%
\bibitem{ars97} A. Armoni and J. Sonnenschein
{\it ``Screening and Confinement in Large $N_f$ $QCD_2$ and in $N=1$ $SYM_2$"}
TAUP-2412-97 hep-th/9703114
%
\bibitem{hak95}
A. Hashimoto and I. R. Klebanov.
{\em Nucl.Phys} {\bf B434}:264-282,1995 hep-th/9409064
%
\bibitem{bfss97} T. Banks, W. Fischler, S. Shenker, L. Susskind,
 {\em Phys. Rev.} {\bf D55} (1997) 5112, hep--th/9610043.
%
\bibitem{dlcqpapers} H.-C. Pauli and S.J.Brodsky,
{\em Phys.Lett.} {\bf D32} (1985) 1993, 2001.

%
\bibitem{alp98a} F.Antonuccio, S.S.Pinsky and O.Lunin,
To appear.
%
\bibitem{suss97} L.Susskind,
 {\it Another Conjecture About Matrix Theory}, hep-th/9704080.
%
\bibitem{anp97} F. Antonuccio, S.S. Pinsky,
{\em Phys.Lett} {\bf B397}:42-50,1997, hep-th/9612021.
%
\bibitem{tay98} W. Taylor,
{\it ``Lectures on D-Branes, Gauge Theory and M(arices)"}
PUPT-1762, Jun 1997. 80pp. Talk given at 2nd Trieste Conference on Duality
in String Theory,
Trieste, Italy, 16-20 Jun 1997.  hep-th/9801182
%
\bibitem{ghk97} D. J. Gross, A. Hashimoto and I. R.Klebanov,
{\it The Spectrum of a Large N Gauge Theory Near Transition From Confinement
to Screening"}
NSF-ITP-97-133, Oct 1997. 16pp.  hep-th/9710240
%
\bibitem{pin97} S. Pinsky,
{\it The Analog of the t'Hooft Pion with Adjoint Fermions"}
Invited talk at New Nonperturbative Methods and Quantization of the Light
Cone, Les Houches,
France, 24 Feb - 7 Mar 1997. hep-th/9705242
%
\bibitem{pin97a} S. Pinsky,
{\em Phys.Rev} {\bf D56}:5040-5049,1997 hep-th/9612073
%
\bibitem{mrp97} G. McCartor, D. G. Robertson and S. Pinsky
 {\em Phys.Rev} {\bf D56}:1035-1049,1997 hep-th/9612083
%
\bibitem{adb97} F.Antonuccio, S.J.Brodsky, S.Dalley,
{\em Phys.Lett.} {\bf B412} (1997) 104-110.


\end{references}
\end{document}